\documentclass[sigconf]{acmart}

\AtBeginDocument{%
  \providecommand\BibTeX{{%
    \normalfont B\kern-0.5em{\scshape i\kern-0.25em b}\kern-0.8em\TeX}}}

\copyrightyear{2020}
\acmYear{2020}
\acmConference[NanoCom '20]{NanoCom '20: ACM International Conference on Nanoscale Computing and Communication}{September 23--25, 2020}{College Park, MD}
\acmBooktitle{NanoCom '20: ACM International Conference on Nanoscale Computing and Communication, September 23--25, 2020, College Park, MD}
\acmPrice{}
\usepackage[color]{changebar}
\def\todoCtd#1{%
    TODO: #1%
    \ifx&#1&...\fi%
    \endgroup
    \cbend
    \relax
}

\usepackage[detect-all]{siunitx}
\DeclareSIUnit{\Bit}{bit}
\DeclareSIUnit{\sample}{Sa}
\usepackage{graphicx}
\usepackage{tikz}
\usepackage[european,americaninductors]{circuitikz}
\usetikzlibrary{patterns}
\usetikzlibrary{external}
\tikzset{external/system call={pdflatex --shell-escape -halt-on-error -interaction=nonstopmode -jobname "\image" "\texsource"}}
\usepackage{pgfplots}
\usepgfplotslibrary{colorbrewer}
\pgfplotsset{
		cycle list/Set1,
		cycle multiindex* list={
			mark list*\nextlist
			Set1\nextlist
		}
	}
\usepackage{overpic}
\usepackage{subcaption}
\usepgfplotslibrary{groupplots}
\begin{document}

\title{Colour-Specific Microfluidic Droplet Detection for Molecular Communication}

\author{Max Bartunik}
\authornote{Both authors contributed equally to this research.}
\email{max.bartunik@fau.de}
\orcid{0000-0002-3033-5798}
\author{Marco Fleischer}
\authornotemark[1]
\email{marco.fleischer@fau.de}
\affiliation{%
  \institution{Institute for Electronics Engineering, Friedrich-Alexander-University Erlangen-Nuernberg (FAU)}
  \streetaddress{Cauerstr. 9}
  \postcode{91058}
  \city{Erlangen}
  \country{Germany}
}

\author{Werner Haselmayr}
\email{werner.haselmayr@jku.at}
\affiliation{%
  \institution{Institute for Communications Engineering and RF-Systems, Johannes Kepler University Linz (JKU)}
  \streetaddress{Altenberger Strasse 69}
  \postcode{4040}
  \city{Linz}
  \country{Austria}}

\author{Jens Kirchner}
\email{jens.kirchner@fau.de}
\affiliation{%
  \institution{Institute for Electronics Engineering, Friedrich-Alexander-University Erlangen-Nuernberg (FAU)}
    \streetaddress{Cauerstr. 9}
  \postcode{91058}
  \city{Erlangen}
  \country{Germany}}

%%\renewcommand{\shortauthors}{Bartunik and Fleischer, et al.}


\begin{abstract}
Droplet-based microfluidic systems are a promising platform for lab-on-a-chip (LoC) applications. These systems can also be used to enhance LoC applications with integrated droplet control information or for data transmission scenarios in the context of molecular communication. For both use-cases the detection and characterisation of droplets in small microfluidic channels is crucial. So far, only complex lab setups with restricted capabilities have been presented as detection devices. We present a new low-cost and portable droplet detector. The device is used to confidently distinguish between individual droplets in a droplet-based microfluidic system. Using on-off keying a 16-bit sequence is successfully transmitted for the first time with such a setup. Furthermore, the devices capabilities to characterise droplets regarding colour and size are demonstrated. Such an application of a spectral sensor in a microfluidic system presents new possibilities, such as colour-coded data transmission or analysis of droplet content.
\end{abstract}





\maketitle

\section{Introduction}
Molecular communication uses biochemical signals or other information carriers in the nanoscale to achieve data transmission in use-cases where either wired or wireless connections are not feasible. This approach has been experimentally investigated in previous work on centimeter- and decimeter-scale (e.g. \cite{Bartunik2019, Unterweger2018, Grebenstein2019, Farsad2013, Bartunik2020}). For smaller geometries, it can be implemented in a microfluidic setup. To this end, we use droplets to transmit data in a microfluidic channel. The theoretical principles of droplet-based communication have been presented in \cite{Leo2013} and a first practical study was conducted in \cite{Haselmayr2019}.

To enable the practical realisation of droplet-based communication the automated generation and detection of droplets is vital. Challenges are posed by the small size of the microfluidic channels, the requirement to not introduce obstacles into the channel and the goal of a compact and portable system. A promising method for droplet generation has recently been introduced in \cite{Hamidovic2020}. In this work we aim to develop a simple, versatile and portable droplet detection device.

Droplets may be detected either by change of conductivity between two electrodes \cite{Niu2007, Elbuken2011, Isgor2014} or with devices based on optical sensors. As the use of electrodes results in a complicated fabrication procedure and inherently necessitates changing the microfluidic channel to accommodate the electrodes, implementation of an optical sensor principle was chosen.

Various optical sensors utilising cameras \cite{deHeij2000, Hamidovic2020, Golberg2014} or fluorescent agents \cite{Tkaczyk2011, Chen2018} were successfully applied. However, these approaches require a complex lab setup or sophisticated image processing.
%\textcolor{red}{F.e. the camera setup presented by \cite{Golberg2014} consists of multiple cameras for different imaging techniques, a robotic stage, lenses, light sources and a housing containing the whole setup resulting in a bulky and convoluted design.}
The aim of this work is the development of a simple and portable droplet detector. Hence, we initially extend the detection principle based on a light-emitting diode (LED) and a photodetector presented in \cite{Srinivasan2003} and \cite{Trivedi2010} to enable droplet-based data transmission. In particular, we eliminate the need to feed optical fibres into the transmission channel, significantly simplifying the detector design and improving its portability as well as facilitating an uncomplicated exchange of the mounted microfluidic chip. This device is further used to, for the first time, transmit a stream of \SI{16}{\Bit} using droplets, which is an important step towards droplet-based data transmission.

In addition to the simple droplet detection using a photodetector, we implement droplet colour detection using a spectral sensor. This enables a more precise analysis of droplets and allows for more sophisticated data transmission schemes. So far only a rudimentary approach to colour detection based on changes in transparency has been presented \cite{Trivedi2010}. Hence, this is the first device allowing for differentiation between a variety of colours in a droplet-based microfluidic system.

Finally, both sensor systems are simultaneously used to determine crucial droplet parameters. In particular, we demonstrate the capability to measure a droplets size inside the transmission channel.

We present our work in three sections. Section \ref{sec:deviceConception} describes the individual components of the sensor devices design. In Section \ref{sec:results} the devices capabilities regarding droplet detection,  colour characterisation, colour intensity variation and determination of droplet length are evaluated. We end the article in Section \ref{sec:conclusion} with a conclusion.
\section{Device Conception}
\label{sec:deviceConception}
The constructed device consists of two separate optical sensors. An analog infrared photodiode to detect the presence of an individual water droplet in an oil carrier medium and a digital 6-channel colour detector to further characterise the droplet. Both sensors were implemented on a common circuit board together with a suitable microcontroller. Furthermore, a measurement setup to mount microfluidic chips was constructed. The design allows a simple exchange of the microfluidic chip.

\subsection{Infrared Sensor}
\label{sec:infraredSensor}
The infrared sensor consists of a light source (LED) that is focused towards the channel (submillimeter-scale) and a photodetector on the opposite side. The result is a light path that gets interrupted when the optical absorption of the material (i.e. ink droplets) inside the microfluidic channel increases. The wavelength of the light source was chosen to be in the infrared spectrum (\SI{940}{\nano\metre}) as this is close to the maximally absorbed wavelength (approx. \SI{970}{\nano\metre}) in water \cite{Workman2007}.

\begin{figure}
		\centering
        \includegraphics[width=\columnwidth]{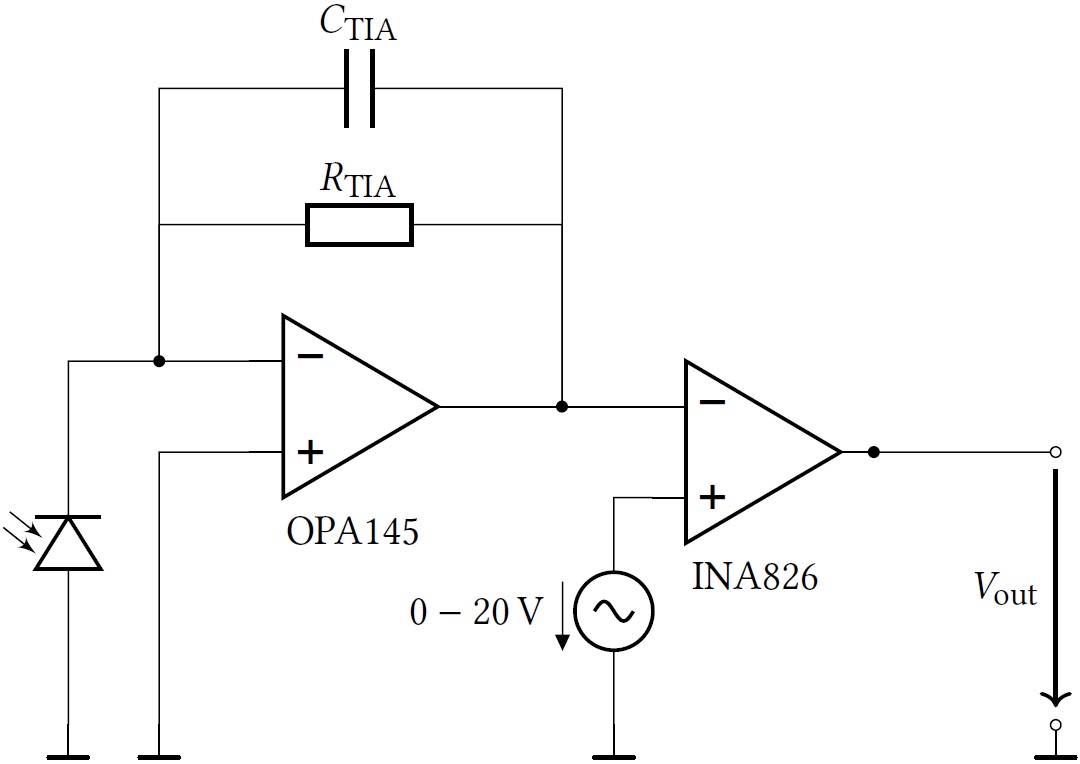}
		\caption{Circuit diagram for the infrared sensor. The current through the photodiode is amplified in two stages (TIA and differential amplifier). The reference voltage at the positive input of the differential amplifier is tunable. $V_\text{out}$ is in the range of \SIrange{0}{5}{\volt}.}
		\label{fig:circuit_irsensor}
\end{figure}

As the signal variation at the photodiode is very small and therefore also the change in diode current, a transimpedance amplifier (TIA) was used to amplify the photodiode output in the range of \SIrange{0}{20}{\volt}. The TIA output is then offset corrected and further amplified using a differential amplifier with a tunable reference voltage. The resulting signal output is adjusted to the range of \SIrange{0}{5}{\volt} and can be acquired with the analog-to-digital converter of a microcontroller. Adequate overvoltage protection at the microcontroller input was implemented. A simplified circuit diagram of the resulting analog setup is shown in Fig. \ref{fig:circuit_irsensor}.

\subsection{Colour Sensor}
To further characterise droplets, a spectral sensor (AS7262) from ams AG was implemented. The sensor has a very compact design and incorporates six photosensors for various wavelengths, covering the range from \SIrange{430}{670}{\nano\metre}. Each photosensor has a bandwidth of \SI{40}{\nano\metre}. As the sensor has integrated digital data acquisition, no analog amplification, as in Fig. \ref{fig:circuit_irsensor} for the infrared sensor, is required.

The recorded spectral data can be retrieved via a standard I\textsuperscript{2}C interface with an appropriate microcontroller. Due to the time required to measure the spectral value and transmit the result with the I\textsuperscript{2}C interface the sample rate is restricted to \SI{20}{\sample\per\second}. A higher sample rate, although possible, would result in reduced sensitivity.
For measurements in the microfluidic setup a white LED was placed opposite to the sensor, analogous to Section \ref{sec:infraredSensor}.

\subsection{PCB and Housing}
The circuitry for both sensors was implemented on a common circuit board, together with a microcontroller (ATmega32U4) from Atmel Corporation, a USB connector and power converters to supply the required voltages for the amplification stages.

The resulting device was constructed as a four layered printed circuit board (PCB), which was designed using Altium Designer.  The PCB was printed by Multi Leiterplatten GmbH in Germany and assembled using the facilities of the Institute for Electronics Engineering at FAU, Germany.

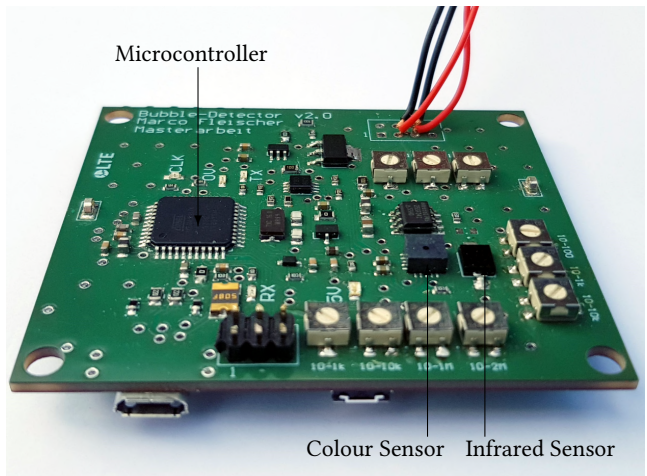
\begin{figure}
   \centering
    \begin{overpic}[width=\columnwidth]{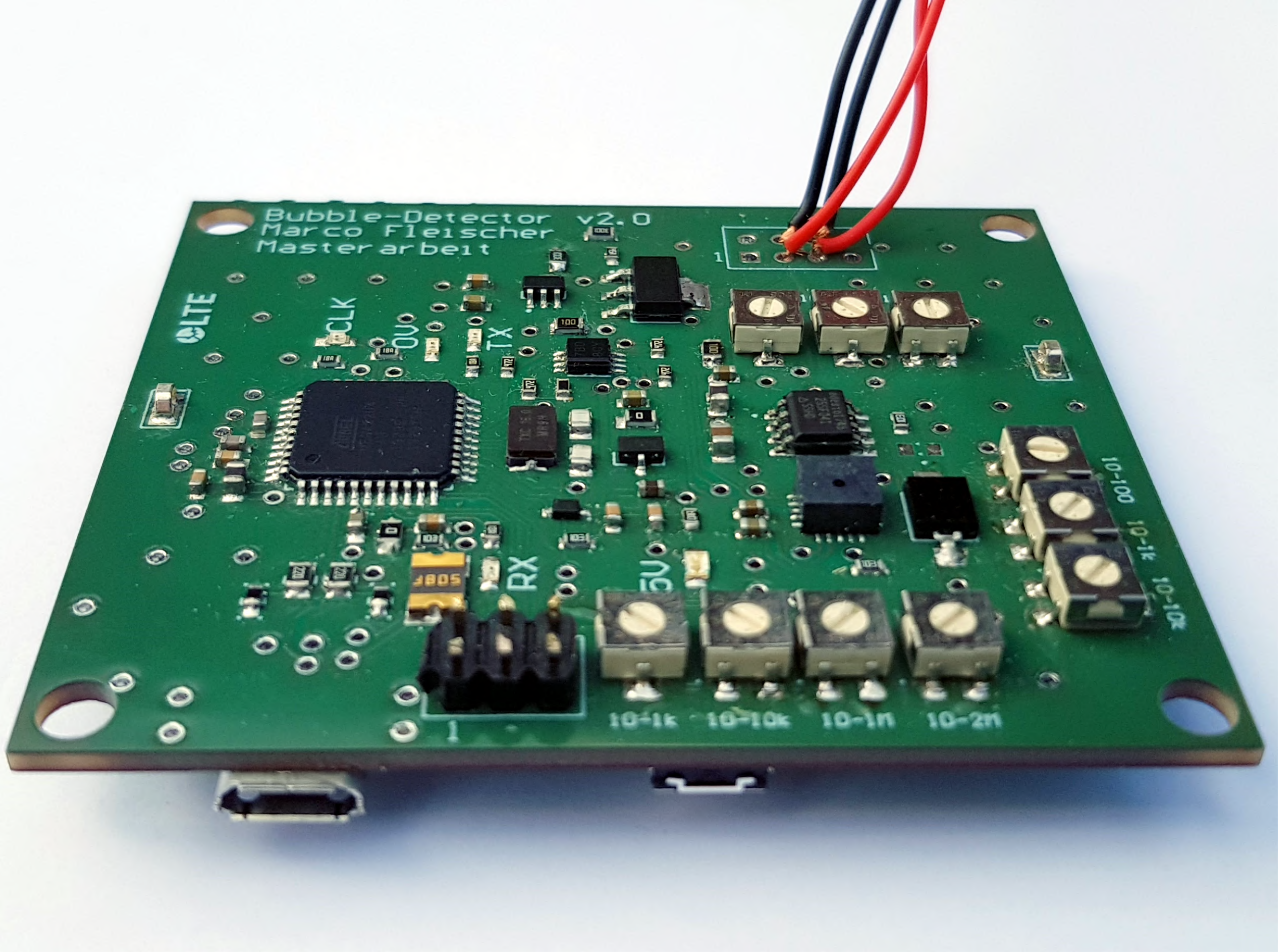}
    \put(17,66){Microcontroller}
    \put(30,65){\vector(0,-1){25}}
    \put(47,4){Colour Sensor}
    \put(66,7){\vector(0,1){27}}
    \put(72,4){Infrared Sensor}
    \put(75,7){\vector(0,1){27}}
    \end{overpic}
    \caption{Assembled PCB with integrated infrared and spectral sensors as well as a microcontroller. A serial connection is provided via a USB port that is also used to power the device. The attached cables connect to the separate LEDs.}
    \label{fig:pcb}
\end{figure}

Fig. \ref{fig:pcb} shows the completed PCB from above. Components were placed on both sides of the circuit board with a size of \SI{6.6}{\centi\metre} by \SI{5.6}{\centi\metre}.

To mount the microfluidic chips for testing a custom-made housing was constructed. This allows for a simple exchange of the used chip and has an adjustable positioning system to accurately place the microfluidic channel on the sensor. A beam across the top of the microfluidic chip acts as a mount for the two light sources.

\begin{figure}
   \centering
    \begin{overpic}[width=0.9\columnwidth]{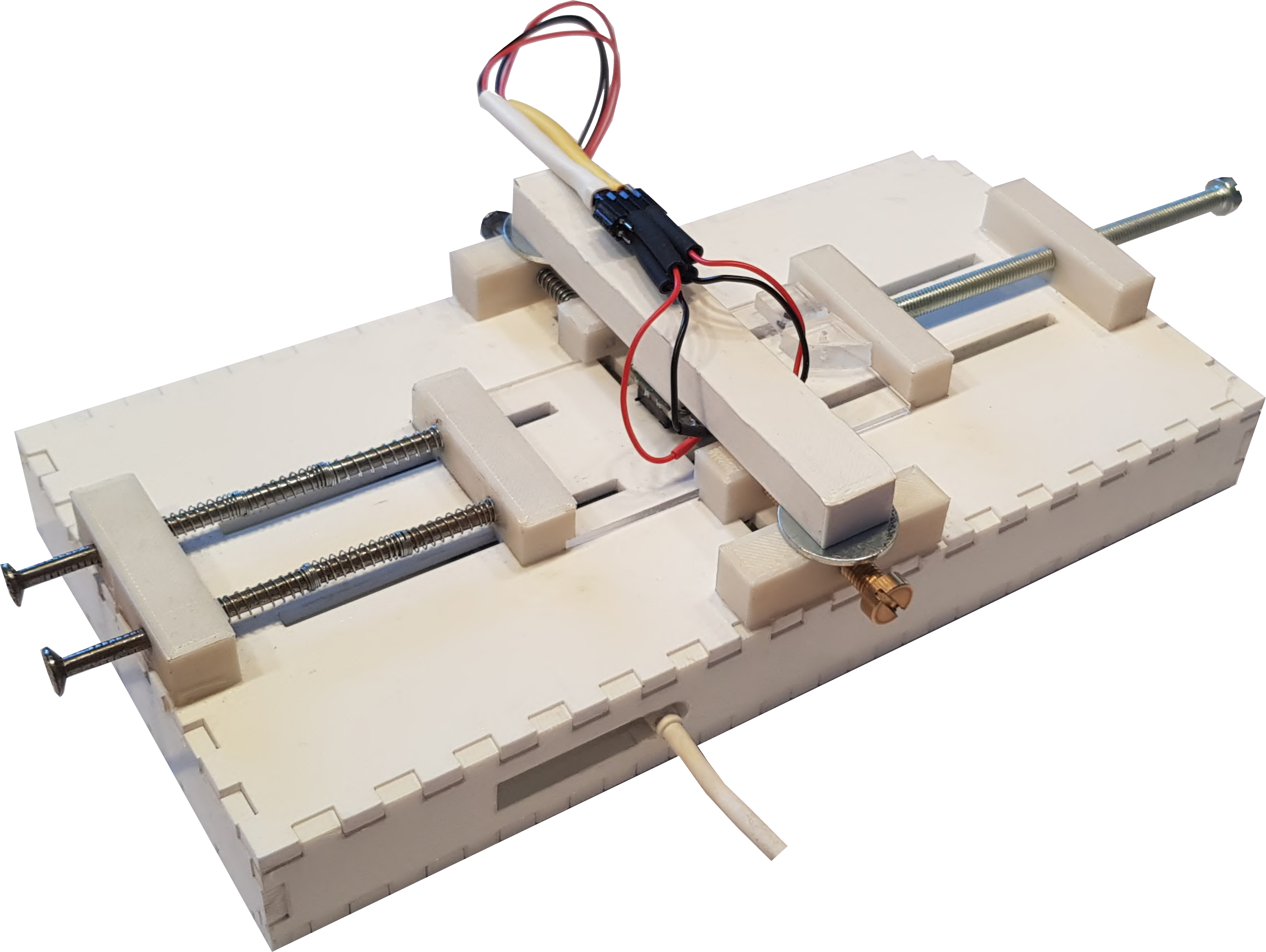}
    \put(0,50){LED-Beam}
    \put(20,51){\vector(1,0){30}}
    \put(70,10){Chip Fixture}
    \put(70,13){\vector(0,1){15}}
    \put(90,13){\vector(0,1){43}}
    %\put(50,0){serial port}
    %\put(60,3){\vector(0,1){5}}
    \put(40,2){PCB-Housing}
    \put(50,7){\vector(0,1){20}}
    \end{overpic}
    \caption{Droplet detection device with positioning system for the microfluidic chip. The light sources (i.e. infrared and white LED) are attached to the beam across the top of the microfluidic chip. The PCB with the sensors is inside the box. The spring-loaded screws can be used to precisely place the microfluidic channel above the sensors.}
    \label{fig:housing}
\end{figure}

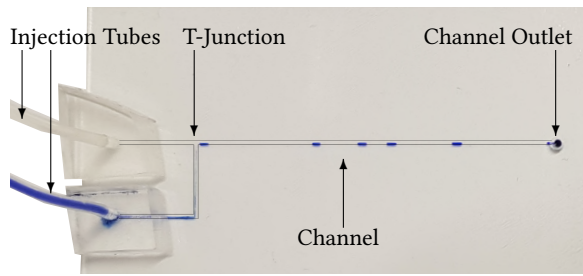
\begin{figure}
   \centering
    \begin{overpic}[width= 0.9\columnwidth, trim={10cm 1cm 1cm 1cm}, clip]{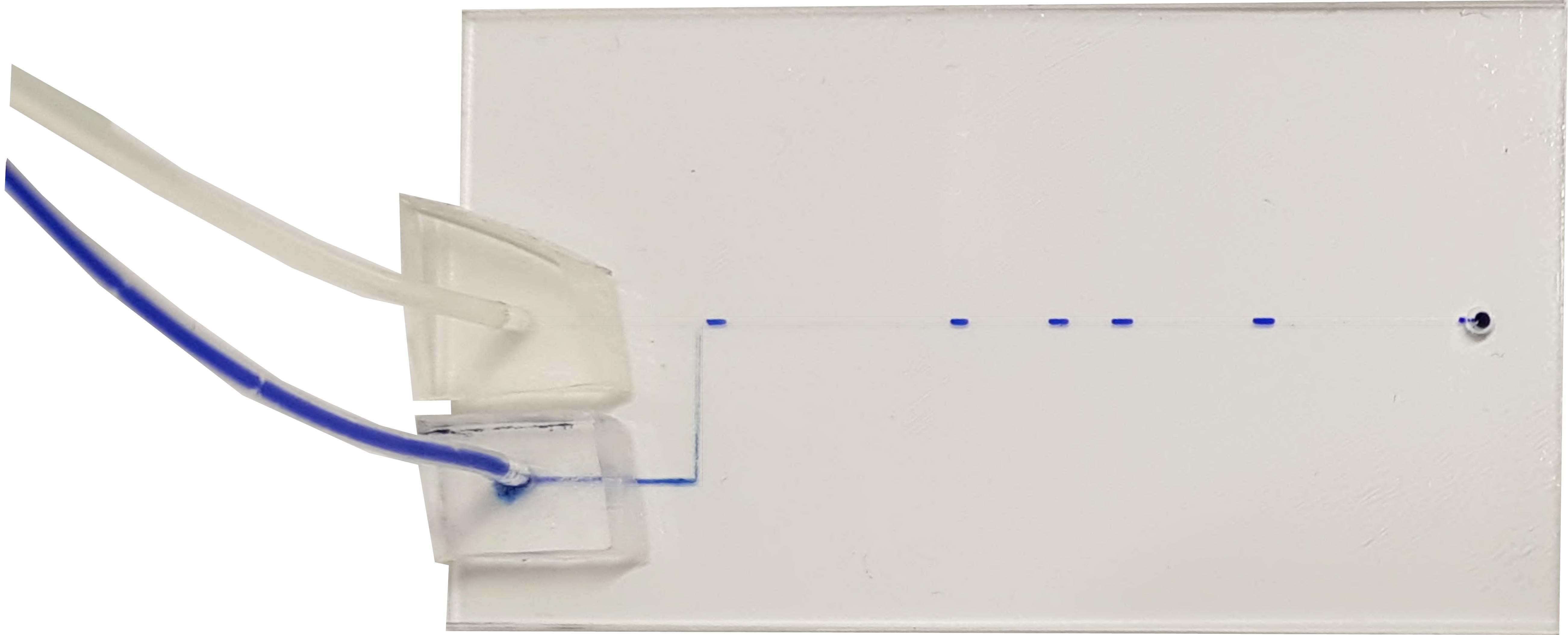}
    \put(50,5){Channel}
    \put(58,8){\vector(0,1){12}}
    \put(72,40){Channel Outlet}
    \put(95,39){\vector(0,-1){15}}
    \put(30,40){T-Junction}
    \put(32,39){\vector(0,-1){15}}
    \put(0,40){Injection Tubes}
    \put(2,39){\vector(0,-1){12}}
    \put(7,39){\vector(0,-1){25}}
    \end{overpic}
    \caption{Microfluidic chip of PMMA with a straight transmission channel and a T-junction used for data transmission. The blue ink is injected at the T-junction creating a droplet stream which can be detected by the proposed device shown in Fig. \ref{fig:pcb}. The channel boundaries were redrawn in the figure for better visibility.}
    \label{fig:bubbles}
\end{figure}

Fig. \ref{fig:housing} shows the complete droplet detection device with a microfluidic chip. Moreover, a microfluidic chip with a stream of inked droplets is shown in Fig. \ref{fig:bubbles}.

\section{Experimental Results}
\label{sec:results}
Various experiments were conducted to evaluate the individual capabilities of the two implemented sensors.
First, the infrared sensor was used as a presence/absence detector for individual droplets. The transmission of a sample data sequence using on-off keying was achieved. Second, the colour sensors spectral sensitivity was investigated using differently coloured droplets. Third, the response of both sensors to various ink concentrations was examined. Finally, the characterisation of droplet size and speed using both sensors was achieved.

The microfluidic chips are made of two overlapping polymethyl methacrylate (PMMA) layers, whereby channels are laser engraved into one of the layers \cite{Haselmayr2019}. In all cases the chip was operated with a pressure controlled pump to inject individual droplets into a constant background flow via a T-junction as proposed in \cite{Hamidovic2020}. This is achieved by increasing the pressure at the injection point relative to a constant pressure in the transmission channel. The background flow consists of regular household oil and the injected droplets of ink diluted with tap water.

\subsection{Data Transmission}
\label{sec:data_transmission}
To achieve data transmission, on-off-keying was implemented. An individual bit was transmitted every second (\SI{1}{\Bit\per\second}), whereas the bit value '1' was coded as the injection of a droplet and the value '0' as no injection. The random 16-bit sequence '10111100 01011001' was used for transmission.

To detect individual bits, the received signal was offset corrected to get a \SI{0}{\volt} baseline and a fixed threshold of \SI{0.2}{V}, significantly higher than the noise level, was applied. Symbol intervals with a known duration were derived from the first threshold pass. A bit was detected as '1' if the threshold was met for at least \SI{30}{\percent} of the symbol interval \footnote{Please note that due to the good signal quality various threshold values lead to a successful data transmission.}. To account for inaccurate symbol intervals due to varying droplet speeds the symbol intervals were re-synchronised at every detected rising edge.

%The good signal quality allows for a variety of parameter settings for the detection threshold.

\begin{figure}
		\centering
        \begin{tikzpicture}
	\begin{axis}[
		xmin = 0, xmax = 20,
		ymin = -0.1, ymax = 1.2,
		xlabel = {Time [\si{\second}]},
		ylabel = {Voltage [\si{\volt}]},
		scaled ticks=false, 
		width = \columnwidth,
		tick label style={/pgf/number format/fixed},
		clip = false
		]
        \begin{scope}
        \clip(axis cs:0,-0.1) rectangle(axis cs: 20,1.2);
        
        \addplot+[mark=none, thick] coordinates {(0,0.2) (20, 0.2)};
        
		\addplot+[mark=none, thick, x filter/.code={\pgfmathparse{\pgfmathresult-16}\pgfmathresult}, y filter/.code={\pgfmathparse{\pgfmathresult-0.34}\pgfmathresult}] table [x=time, y=voltage, col sep=comma] {figures/data/DataTr.csv};

        \addplot[mark=none, black, dashed] coordinates {(1.4,1.2) (1.4,-0.1)};
        \addplot[mark=none, black, dashed] coordinates {(2.4,1.2) (2.4,-0.1)};
        \addplot[mark=none, black, dashed] coordinates {(3.5,1.2) (3.5,-0.1)};
        %\addplot[mark=none, black, dashed] coordinates {(3.43,2) (3.43,-0.1)};
        \addplot[mark=none, black, dashed] coordinates {(4.5,1.2) (4.5,-0.1)};
        \addplot[mark=none, black, dashed] coordinates {(5.7,1.2) (5.7,-0.1)};
        \addplot[mark=none, black, dashed] coordinates {(6.75,1.2) (6.75,-0.1)};
        %\addplot[mark=none, black, dashed] coordinates {(6.9,2) (6.9,-0.1)};
        \addplot[mark=none, black, dashed] coordinates {(7.75,1.2) (7.75,-0.1)};
        \addplot[mark=none, black, dashed] coordinates {(8.75,1.2) (8.75,-0.1)};
        \addplot[mark=none, black, dashed] coordinates {(9.75,1.2) (9.75,-0.1)};
        \addplot[mark=none, black, dashed] coordinates {(10.9,1.2) (10.9,-0.1)};
        \addplot[mark=none, black, dashed] coordinates {(11.9,1.2) (11.9,-0.1)};
        \addplot[mark=none, black, dashed] coordinates {(12.9,1.2) (12.9,-0.1)};
        %\addplot[mark=none, black, dashed] coordinates {(13.2,2) (13.2,-0.1)};
        \addplot[mark=none, black, dashed] coordinates {(14.2,1.2) (14.2,-0.1)};
        \addplot[mark=none, black, dashed] coordinates {(15.2,1.2) (15.2,-0.1)};
        \addplot[mark=none, black, dashed] coordinates {(16.2,1.2) (16.2,-0.1)};
        \addplot[mark=none, black, dashed] coordinates {(17.2,1.2) (17.2,-0.1)};
        %\addplot[mark=none, black, dashed] coordinates {(17.5,2) (17.5,-0.1)};
        \addplot[mark=none, black, dashed] coordinates {(18.2,1.2) (18.2,-0.1)};

	    \end{scope}
	    
	    \node at (axis cs: 1.9 , 1.2)[above] {1};
	    \node at (axis cs: 2.9 , 1.2)[above] {0};
	    \node at (axis cs: 4   , 1.2)[above] {1};
	    \node at (axis cs: 5   , 1.2)[above] {1};
	    \node at (axis cs: 6.2 , 1.2)[above] {1};
	    \node at (axis cs: 7.25, 1.2)[above] {1};
	    \node at (axis cs: 8.25, 1.2)[above] {0};
	    \node at (axis cs: 9.25, 1.2)[above] {0};
	    \node at (axis cs: 10.25, 1.2)[above] {0};
	    \node at (axis cs: 11.4 , 1.2)[above] {1};
	    \node at (axis cs: 12.4 , 1.2)[above] {0};
	    \node at (axis cs: 13.55, 1.2)[above] {1};
	    \node at (axis cs: 14.7 , 1.2)[above] {1};
	    \node at (axis cs: 15.7 , 1.2)[above] {0};
	    \node at (axis cs: 16.7 , 1.2)[above] {0};
	    \node at (axis cs: 17.85, 1.2)[above] {1};
	    
    \end{axis}
		
\end{tikzpicture}
		\caption{Receive signal for the transmitted bit sequence '10111100 01011001' (shown in blue) together with the signal threshold at \SI{0.2}{\volt} (shown in red) and the individual symbol intervals (dashed lines). A symbol was detected as '1' if the threshold was met for at least \SI{30}{\percent} of a symbol interval. All 16 bits of the sample sequence were decoded correctly.}
		\label{fig:transmission_infrared}
\end{figure}
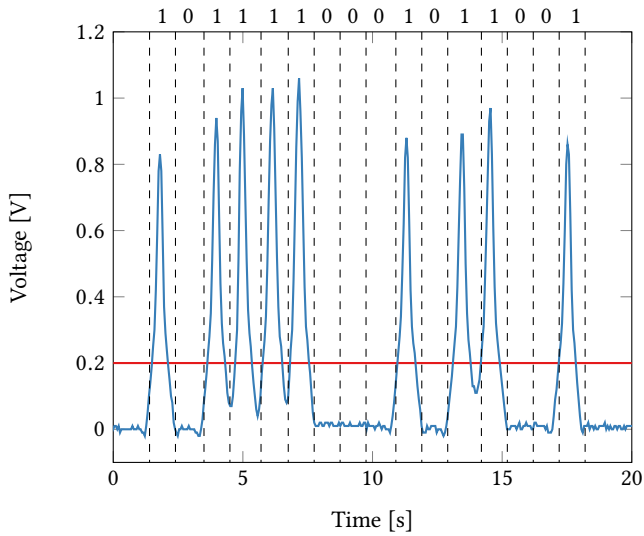

Fig. \ref{fig:transmission_infrared} shows the receive signal for the transmitted sample bit sequence. The detection threshold and the relevant symbol intervals, as derived from the the rising edges, are also shown. As can be seen each transmitted bit was successfully received with the device and the scheme described above.

\subsection{Droplet Colour Determination}
The spectral sensors main function is to determine the colour of a detected droplet. However, it can also be used as an extension to the analog infrared sensor. Possible use-cases for colour detection are more sophisticated data transmission schemes and the characterisation of droplet content (e.g. with a coloured reagent) in LoC applications.

To assess the colour-coded droplet detection the measured spectral signal for ink droplet of various colours was recorded. Fig. \ref{fig:droplet_colour} shows the normalised sensor values for the individual droplets. One droplet for each of the six different ink colours (violet, blue, green, yellow, orange and red) was transmitted. Significantly different colours can clearly be identified due to their spectral values (e.g. blue and violet) using the constructed sensor. Spectral values for related colours (e.g. yellow and orange) are very similar and require further investigation to ensure distinct detection. An in-depth analysis of colour sensitivity may be achieved by investigating the sensor data for gradual transitions between colours with a series of dilutions.

\begin{figure}
		\centering
        \begin{tikzpicture}
	\begin{axis}[
		xmin = 5, xmax = 25,
		ymin = 0, ymax = 1.1,
		xlabel = {Time [\si{\second}]},
		ylabel = {Normalised Intensity},
        legend entries = {\SI{450}{\nano\meter}, \SI{500}{\nano\meter}, \SI{550}{\nano\meter}, \SI{570}{\nano\meter}, \SI{600}{\nano\meter}, \SI{650}{\nano\meter}},
    	legend columns = 3,
    	legend pos = {south west},
		width = \columnwidth,
		tick label style={/pgf/number format/fixed},
		grid = both,
		minor tick num = 4
		]
		%pgfmath seems to have a problem with legend entries - solved with alternative math definition
		\addplot+[Set1-D, mark=none, x filter/.code={\pgfmathparse{\pgfmathresult-8.5}\pgfmathresult}] table [x=time, y=violet, col sep=comma] {figures/data/DataTr_colour.csv};
		\addplot+[Set1-B, mark=none, x filter/.code={\pgfmathparse{\pgfmathresult-8.5}\pgfmathresult}] table [x=time, y=blue, col sep=comma] {figures/data/DataTr_colour.csv};
		\addplot+[Set1-C, mark=none, x filter/.code={\pgfmathparse{\pgfmathresult-8.5}\pgfmathresult}] table [x=time, y=green, col sep=comma] {figures/data/DataTr_colour.csv};
		\addplot+[Set1-F, mark=none, x filter/.code={\pgfmathparse{\pgfmathresult-8.5}\pgfmathresult}] table [x=time, y=yellow, col sep=comma] {figures/data/DataTr_colour.csv};
		\addplot+[Set1-E, mark=none, x filter/.code={\pgfmathparse{\pgfmathresult-8.5}\pgfmathresult}] table [x=time, y=orange, col sep=comma] {figures/data/DataTr_colour.csv};
		\addplot+[Set1-A, mark=none, x filter/.code={\pgfmathparse{\pgfmathresult-8.5}\pgfmathresult}] table [x=time, y=red, col sep=comma] {figures/data/DataTr_colour.csv};

    \end{axis}
\end{tikzpicture}
		\caption{Recorded spectral values for the sample transmission using red ink in Fig. \ref{fig:transmission_infrared}. Each '1' bit can be seen as an individual notch. The colour values are consistent for the whole transmission.}
		\label{fig:transmission_colour}
\end{figure}
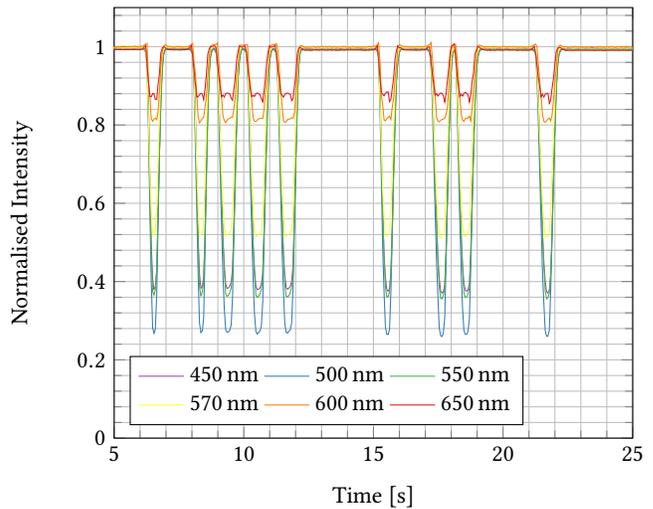

Fig. \ref{fig:transmission_colour} shows the spectral values that were recorded by the colour sensor for the transmission of the 16-bit sequence introduced in Sec. \ref{sec:data_transmission}. The data points are set at the center wavelength of the six photosensors.

Each bit can clearly be identified with an equal colour distribution equivalent to a red droplet in Fig. \ref{fig:droplet_colour}. The sensor can therefore consistently characterise droplet colour throughout a transmission sequence, allowing for colour coded data transmission by differentiation between droplet spectra (as shown in Fig. \ref{fig:droplet_colour}). 

\begin{figure}
	\centering
		\begin{tikzpicture}
    		\begin{axis}[
    		xlabel = {Wavelength [\si{\nano\metre}]},
    		ylabel = {Normalised Intensity},
    		ymax = 1,
    		ymin = 0,
    		legend entries = {Red, Blue, Green, Violet, Orange, Yellow},
    		legend columns = 3,
    		legend pos = {south west},
    		grid = both,
    		width = \columnwidth
    		]
    		\addplot  table [x=wv, y=int, col sep=comma]{figures/data/Red_Droplet.csv};
    		\addplot  table [x=wv, y=int, col sep=comma]{figures/data/Blue_Droplet.csv};
    		\addplot  table [x=wv, y=int, col sep=comma]{figures/data/Green_Droplet.csv};
    		\addplot  table [x=wv, y=int, col sep=comma]{figures/data/Violet_Droplet.csv};
    		\addplot  table [x=wv, y=int, col sep=comma]{figures/data/Orange_Droplet.csv};
    		\addplot  table [x=wv, y=int, col sep=comma]{figures/data/Yellow_Droplet.csv};
    		\end{axis}
\end{tikzpicture}
	\caption{Spectral comparison of various droplet colours using the constructed colour sensor. The data is normalised to the spectral profile recorded without droplets. Colours can clearly be differentiated.}
	\label{fig:droplet_colour}
\end{figure}
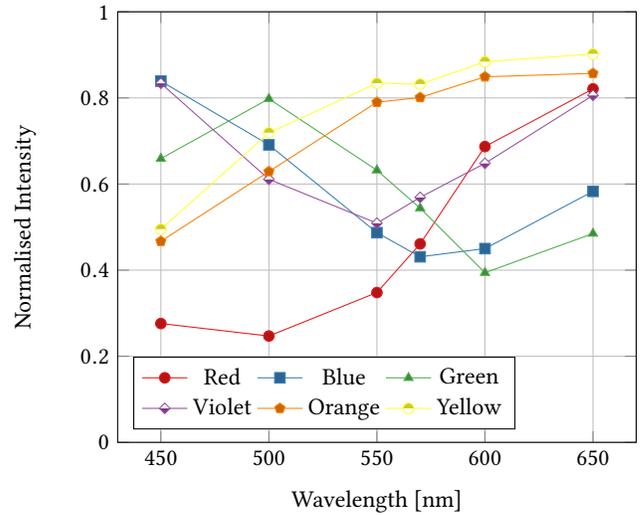

\subsection{Variation of Droplet Concentration}
In order to assess the sensitivity of the sensors, a series of dilutions with varying ink concentrations was performed. Furthermore, the infrared sensors operation based on absorption in water (independent of colouring) is demonstrated with clear water droplets.

Water was diluted with blue ink in concentrations ranging from \SIrange{0}{25}{\percent} in steps of \SI{5}{\percent}. For each measurement a droplet of \SI{5}{\milli\metre} length was generated.

\begin{figure}
		\centering
        \begin{tikzpicture}
  \begin{groupplot}[group style={group size=6 by 1, 
                                ylabels at=edge left,
                                yticklabels at=edge left,
                                xlabels at=edge bottom,
                                xticklabels at=edge bottom,
                                horizontal sep=2pt},
                                height = 8cm,
                                width=.31\columnwidth,
                                tickpos=left,
                                xmin = -0.5,
                                xmax = 6.5,
                                ymax = 2,
                                ymin = -0.1,
                                ytick align=outside,
                                xtick align=outside,
                                xtick = {0,1,2,3,4,5,6},
                                xticklabels = {0,,,,4,,},
		                        ylabel = {Voltage [\si{\volt}]}]
		                        
    \nextgroupplot[title=\SI{0}{\percent}]
        \addplot+[mark=none, thick] coordinates {(-0.5,0.7275) (6.5, 0.7275)};
    	\addplot+[mark=none, thick] table [x=time, y=voltage, col sep=comma] {figures/data/Water.csv};
    	\addplot+[mark=none, thick] coordinates {(-0.5,0.667) (6.5, 0.667)};
    \nextgroupplot[title= \SI{5}{\percent}]
        \addplot+[mark=none, thick] coordinates {(-0.5,0.7275) (6.5, 0.7275)};
        \addplot+[mark=none, thick] table [x=time, y=voltage, col sep=comma] {figures/data/0,5.csv};
        \addplot+[mark=none, thick] coordinates {(-0.5,0.857) (6.5, 0.857)};
    \nextgroupplot[title= \SI{10}{\percent}, xlabel = {Time [\si{\second}]},every axis x label/.append style={at=(ticklabel cs:1.1)}]
        \addplot+[mark=none, thick] coordinates {(-0.5,0.7275) (6.5, 0.7275)};
        \addplot+[mark=none, thick] table [x=time, y=voltage, col sep=comma] {figures/data/1,0.csv};
        \addplot+[mark=none, thick] coordinates {(-0.5,0.753) (6.5, 0.753)};
    \nextgroupplot[title= \SI{15}{\percent}]
        \addplot+[mark=none, thick] coordinates {(-0.5,0.7275) (6.5, 0.7275)};
        \addplot+[mark=none, thick] table [x=time, y=voltage, col sep=comma] {figures/data/1,5.csv};
        \addplot+[mark=none, thick] coordinates {(-0.5,0.758) (6.5, 0.758)};
    \nextgroupplot[title=\SI{20}{\percent}]
        \addplot+[mark=none, thick] coordinates {(-0.5,0.7275) (6.5, 0.7275)};
        \addplot+[mark=none, thick] table [x=time, y=voltage, col sep=comma] {figures/data/2,0.csv};
        \addplot+[mark=none, thick] coordinates {(-0.5,0.63) (6.5, 0.63)};
    \nextgroupplot[title=\SI{25}{\percent}]
        \addplot+[mark=none, thick] coordinates {(-0.5,0.7275) (6.5, 0.7275)};
        \addplot+[mark=none, thick] table [x=time, y=voltage, col sep=comma] {figures/data/2,5.csv};
        \addplot+[mark=none, thick] coordinates {(-0.5,0.7) (6.5, 0.7)};
  \end{groupplot}
\end{tikzpicture}
		\caption{Measurements of water droplets with varying concentrations of blue ink in the range of \SIrange{0}{25}{\percent} using the infrared sensor (shown in blue). The mean value for each droplet (shown in green) as well as the overall average (shown in red) was calculated. A similar signal can be observed in all cases, showing the specificity to absorption in water independent of colouring }
		\label{fig:transparency_infrared}
\end{figure}
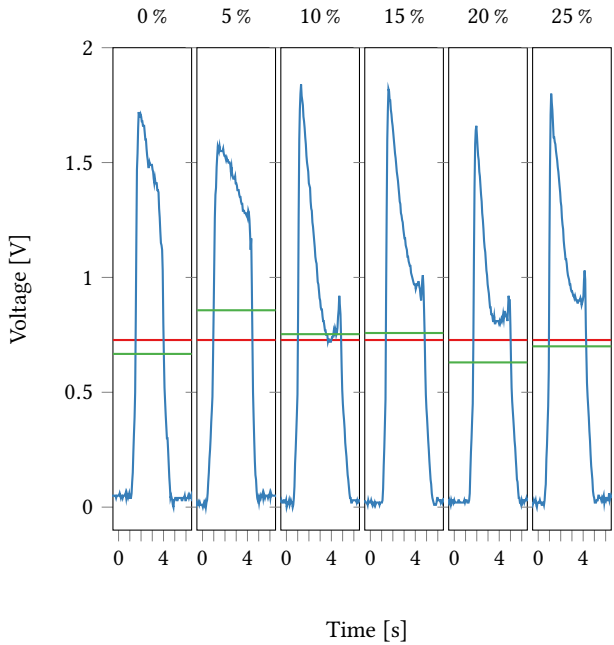

As can be seen in Fig. \ref{fig:transparency_infrared}, the measured signal amplitude of the infrared sensor is constant for various ink concentrations. The calculated mean signal is \SI{0.73}{\volt} with a maximal deviation of \SI{0.13}{\volt} from the average value for the individual concentrations. As expected, the sensors receive signal is independent of colouring as it relies on the absorption in water.

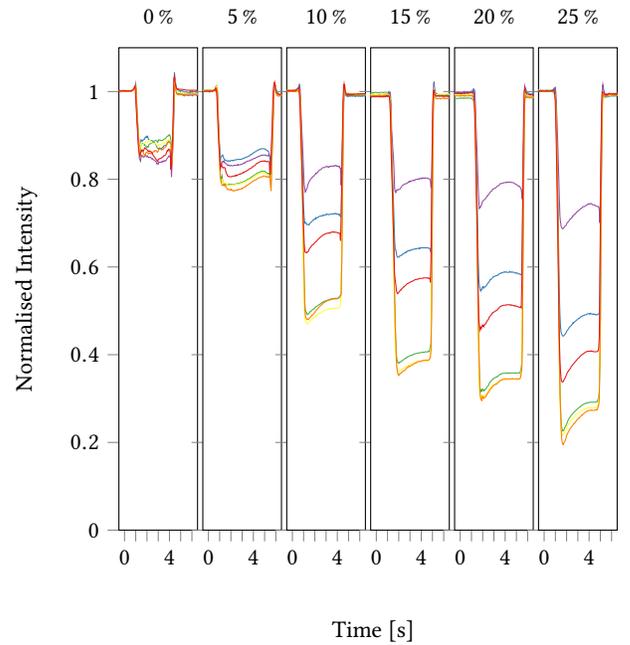
\begin{figure}
		\centering
        \begin{tikzpicture}
  \begin{groupplot}[group style={group size=6 by 1, 
                                ylabels at=edge left,
                                yticklabels at=edge left,
                                xlabels at=edge bottom,
                                xticklabels at=edge bottom,
                                horizontal sep=2pt},
                                height = 8cm,
                                width=.31\columnwidth,
                                tickpos=left,
                                xmin = -0.5,
                                xmax = 6.5,
                                ymax = 1.1,
                                ymin = 0,
                                ytick align=outside,
                                xtick align=outside,
                                xtick = {0,1,2,3,4,5,6},
                                xticklabels = {0,,,,4,,},
		                        ylabel = {Normalised Intensity}]
    \nextgroupplot[title= \SI{0}{\percent}]
        \addplot+[Set1-D, mark=none] table [x=time, y=violet, col sep=comma] {figures/data/Water_colour.csv};
		\addplot+[Set1-B, mark=none] table [x=time, y=blue, col sep=comma] {figures/data/Water_colour.csv};
		\addplot+[Set1-C, mark=none] table [x=time, y=green, col sep=comma] {figures/data/Water_colour.csv};
		\addplot+[Set1-F, mark=none] table [x=time, y=yellow, col sep=comma] {figures/data/Water_colour.csv};
		\addplot+[Set1-E, mark=none] table [x=time, y=orange, col sep=comma] {figures/data/Water_colour.csv};
		\addplot+[Set1-A, mark=none] table [x=time, y=red, col sep=comma] {figures/data/Water_colour.csv};

    \nextgroupplot[title= \SI{5}{\percent}]
        \addplot+[Set1-D, mark=none] table [x=time, y=violet, col sep=comma] {figures/data/0,5_colour.csv};
		\addplot+[Set1-B, mark=none] table [x=time, y=blue, col sep=comma] {figures/data/0,5_colour.csv};
		\addplot+[Set1-C, mark=none] table [x=time, y=green, col sep=comma] {figures/data/0,5_colour.csv};
		\addplot+[Set1-F, mark=none] table [x=time, y=yellow, col sep=comma] {figures/data/0,5_colour.csv};
		\addplot+[Set1-E, mark=none] table [x=time, y=orange, col sep=comma] {figures/data/0,5_colour.csv};
		\addplot+[Set1-A, mark=none] table [x=time, y=red, col sep=comma] {figures/data/0,5_colour.csv};
    
    \nextgroupplot[title= \SI{10}{\percent}, xlabel = {Time [\si{\second}]},every axis x label/.append style={at=(ticklabel cs:1.1)}]
        \addplot+[Set1-D, mark=none] table [x=time, y=violet, col sep=comma] {figures/data/1,0_colour.csv};
		\addplot+[Set1-B, mark=none] table [x=time, y=blue, col sep=comma] {figures/data/1,0_colour.csv};
		\addplot+[Set1-C, mark=none] table [x=time, y=green, col sep=comma] {figures/data/1,0_colour.csv};
		\addplot+[Set1-F, mark=none] table [x=time, y=yellow, col sep=comma] {figures/data/1,0_colour.csv};
		\addplot+[Set1-E, mark=none] table [x=time, y=orange, col sep=comma] {figures/data/1,0_colour.csv};
		\addplot+[Set1-A, mark=none] table [x=time, y=red, col sep=comma] {figures/data/1,0_colour.csv};
		
 \nextgroupplot[title= \SI{15}{\percent}]
        \addplot+[Set1-D, mark=none] table [x=time, y=violet, col sep=comma] {figures/data/1,5_colour.csv};
		\addplot+[Set1-B, mark=none] table [x=time, y=blue, col sep=comma] {figures/data/1,5_colour.csv};
		\addplot+[Set1-C, mark=none] table [x=time, y=green, col sep=comma] {figures/data/1,5_colour.csv};
		\addplot+[Set1-F, mark=none] table [x=time, y=yellow, col sep=comma] {figures/data/1,5_colour.csv};
		\addplot+[Set1-E, mark=none] table [x=time, y=orange, col sep=comma] {figures/data/1,5_colour.csv};
		\addplot+[Set1-A, mark=none] table [x=time, y=red, col sep=comma] {figures/data/1,5_colour.csv};

	\nextgroupplot[title= \SI{20}{\percent}]
        \addplot+[Set1-D, mark=none] table [x=time, y=violet, col sep=comma] {figures/data/2,0_colour.csv};
		\addplot+[Set1-B, mark=none] table [x=time, y=blue, col sep=comma] {figures/data/2,0_colour.csv};
		\addplot+[Set1-C, mark=none] table [x=time, y=green, col sep=comma] {figures/data/2,0_colour.csv};
		\addplot+[Set1-F, mark=none] table [x=time, y=yellow, col sep=comma] {figures/data/2,0_colour.csv};
		\addplot+[Set1-E, mark=none] table [x=time, y=orange, col sep=comma] {figures/data/2,0_colour.csv};
		\addplot+[Set1-A, mark=none] table [x=time, y=red, col sep=comma] {figures/data/2,0_colour.csv};

    \nextgroupplot[title= \SI{25}{\percent}]
        \addplot+[Set1-D, mark=none] table [x=time, y=violet, col sep=comma] {figures/data/2,5_colour.csv};
		\addplot+[Set1-B, mark=none] table [x=time, y=blue, col sep=comma] {figures/data/2,5_colour.csv};
		\addplot+[Set1-C, mark=none] table [x=time, y=green, col sep=comma] {figures/data/2,5_colour.csv};
		\addplot+[Set1-F, mark=none] table [x=time, y=yellow, col sep=comma] {figures/data/2,5_colour.csv};
		\addplot+[Set1-E, mark=none] table [x=time, y=orange, col sep=comma] {figures/data/2,5_colour.csv};
		\addplot+[Set1-A, mark=none] table [x=time, y=red, col sep=comma] {figures/data/2,5_colour.csv};

  \end{groupplot}
\end{tikzpicture}
		\caption{Recorded signal of the spectral sensor for \SI{5}{\milli\metre} long droplets with varying concentrations of blue ink. The observed sensor intensity decreases, as light absorption increases, proportional to the amount of ink in the droplet.}
		\label{fig:transparency_colour}
\end{figure}

Fig. \ref{fig:transparency_colour} shows the measured signal for different ink concentrations using the colour sensor. As the ink concentration increases a more distinct colour distribution can be observed. The light absorption increases with the used amount of ink.

The measurements show, that the sensor system can be used to distinguish between different colour concentrations (i.e. colour intensities). Furthermore, the infrared sensor can be used for presence/absence detection of water droplets, independent of colouring. In future applications this may be used to differentiate between different droplet substances.

\subsection{Droplet Size Determination}
The sensor can also be used to measure the size of an individual droplet by observing the infrared and colour sensor values on a time scale. The observed duration of a droplet is proportional to the travel speed inside the microfluidic channel and the volume of the droplet. Considering the speed of the droplet in the channel $v_\text{chan}$ the droplet length can be calculated from the observed droplet duration $t_\text{drop}$ as
\begin{equation}
    l_\text{drop} = v_\text{chan} t_\text{drop} = \frac{\Delta d}{\Delta t} t_\text{drop}
\end{equation}
where $\Delta d = \SI{6.35}{\milli\metre}$ is the distance (fixed on the PCB) and $\Delta t$ the time delay of the received signal edge between the two sensors. Either of the sensors can be used to measure the droplet duration. In this case, the colour sensor was chosen and $t_\text{drop}$ was set to the time span for which the recorded signal surpassed half of the maximal amplitude (full width at half maximum).

\begin{figure}
		\centering
        \begin{tikzpicture}
	\begin{axis}[
		xmin = 0.5, xmax = 5.5,
		ymin = 0.5, ymax = 5.5,
		xlabel = {Expected Length [\si{\milli\metre}]},
		ylabel = {Measured Length [\si{\milli\metre}]},
		scaled ticks=false, 
		width = \columnwidth,
		tick label style={/pgf/number format/fixed},
		ymajorgrids = true
		]
    
        \addplot+[Set1-B, only marks] plot [error bars/.cd, y dir=both, y explicit] table [x = x, y = y ,y error plus=ymax, y error minus=ymin, col sep =comma] {figures/data/dropletSizeErrors_Deviation.csv};
        
    \end{axis}
		
\end{tikzpicture}
		\caption{Measurements to evaluate precision of droplet size values. The average determined droplet size and standard deviation for droplets ranging from \SIrange{1}{5}{\milli\metre} is shown.}
		\label{fig:dropletSize}
\end{figure}
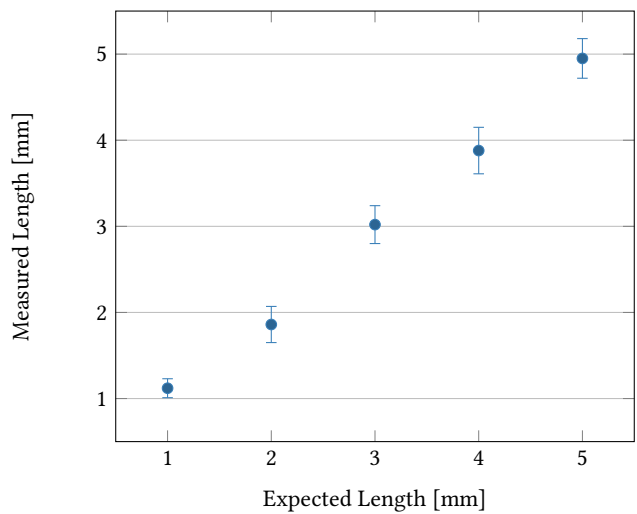

Droplets of varying length (\SIrange{1}{5}{\milli\metre}) were generated by adjusting the duration of a rectangular pressure pulse from \SIrange{1}{5}{\second}, to evaluate the sensors ability to determine droplet size. Fig. \ref{fig:dropletSize} shows the measured droplet length in relation to the transmitted droplet length. Five different droplet lengths were investigated with 20 measurements per value. We observed a precision of \SI{0.48}{\milli\metre}, corresponding to the largest deviation from the expected value.

As the tested droplet lengths can be differentiated, it is possible to transmit data encoded as droplet size in the microfluidic system to achieve higher bit-rates \cite{Leo2013}. The observed precision not only depends on the receiver but also on the reproducible accuracy of the transmitter. A higher precision, resulting in the possibility to encode data in a smaller size variation, might therefore be observed with an improved droplet generation method.

\section{Conclusion}
\label{sec:conclusion}
A low-cost and portable sensor for droplets in microfluidic systems was constructed. The device was successfully used for presence/absence detection and to characterise droplet colour and size. Furthermore, a distinction between different ink concentrations was achieved. It therefore facilitates LoC use cases that require sophisticated data transmission schemes or colour reagent detection.

With the current setup a bit rate of approximately \SI{1}{\Bit\per\second} was achieved. Limiting factors can be found both in parameters of droplet generation (precision of volume and timing) and detection (local resolution and sample rate). The precision of droplet generation may be improved with other pump systems, such as a micropump or a peristaltic pump. Furthermore, implementing a feedback loop to control a pressure pump, based on the retrieved sensor data, may improve droplet generation consistency. If necessary, the sample rate of the sensor device (currently \SI{20}{\sample\per\second}) could be increased by use of an array with multiple sensors in parallel. By addressing these limitations the bit rate could be increased further in an optimised setup.
 
In future work various functions of the sensor device may be extended. Specifically, for the data transmission scenario, the bit rate could be increased by combining coding schemes such as coding of droplet position, colour and size. Furthermore, the development of a standardised interface for microfluidic chips could simplify interoperability for future devices. Such a standard could include dedicated positions for in- and output connectors, sensing regions and a scalable set of microfluidic chip dimensions. Finally, the sensors capabilities regarding sensitivity and more complex coding schemes will be investigated further.

\begin{acks}
The authors would like to express their sincere gratitude to Dominik Lehner for his
support with the microfluidic setup.
\end{acks}

\bibliographystyle{ACM-Reference-Format}
\bibliography{bibliography}

%\appendix

\end{document}